\documentclass[
    ,final            
  ]
  {aipproc}

\layoutstyle{6x9}

\newcommand{\arcsec}{^{\prime\prime}}
\newcommand{\ks}{K_{\rm S}}
\newcommand{\aap}{A\&A}

\newcommand{\apj}{ApJ}
\newcommand{\apjl}{ApJL}

\newcommand{\mnras}{MNRAS}
\newcommand{\nat}{Nat}

\usepackage{subfigure}

\begin{document}

\title{Galactic Centre X-ray Sources}


\author{A.~J. Gosling}{
  address={Department of Astrophysics, University of Oxford, Oxford, OX1 3RH, U.K.}
}

\author{R.~M. Bandyopadhyay}{
  address={Department of Astrophysics, University of Florida, Florida, U.S.A.}
}

\author{K.~M. Blundell}{
  address={Department of Astrophysics, University of Oxford, Oxford, OX1 3RH, U.K.}
}

\begin{abstract}

We report on a campaign to identify the counterparts to the population
of X-ray sources discovered at the centre of our Galaxy by
\citet{wang02} using {\it Chandra}. We have used deep, near infrared
images obtained on VLT/ISAAC to identify candidate counterparts as
astrometric matches to the X-ray positions. Follow up $\ks$-band
spectroscopic observations of the candidate counterparts are used to
search for accretions signatures in the spectrum, namely the
Brackett-$\gamma$ emission line \citep{band97}. From our small initial
sample, it appears that only a small percentage, $\sim 2 - 3 \%$ of
the $\sim 1000$ X-ray sources are high mass X-ray binaries or wind
accreting neutron stars, and that the vast majority will be shown to
be canonical low mass X-ray binaries and cataclysmic variables.

\end{abstract}

\maketitle

\section{Search for counterparts}

Imaging 26 fields towards the Galactic Centre in the near infrared
with VLT/ISAAC, we were able to astrometrically identify possible
counterparts to 79 of the $\sim 1000$ weak, hard X-ray sources
discovered in the {\it Chandra} survey of \citet{wang02}.

\begin{figure}[h]
  \includegraphics[height=.2\textheight]{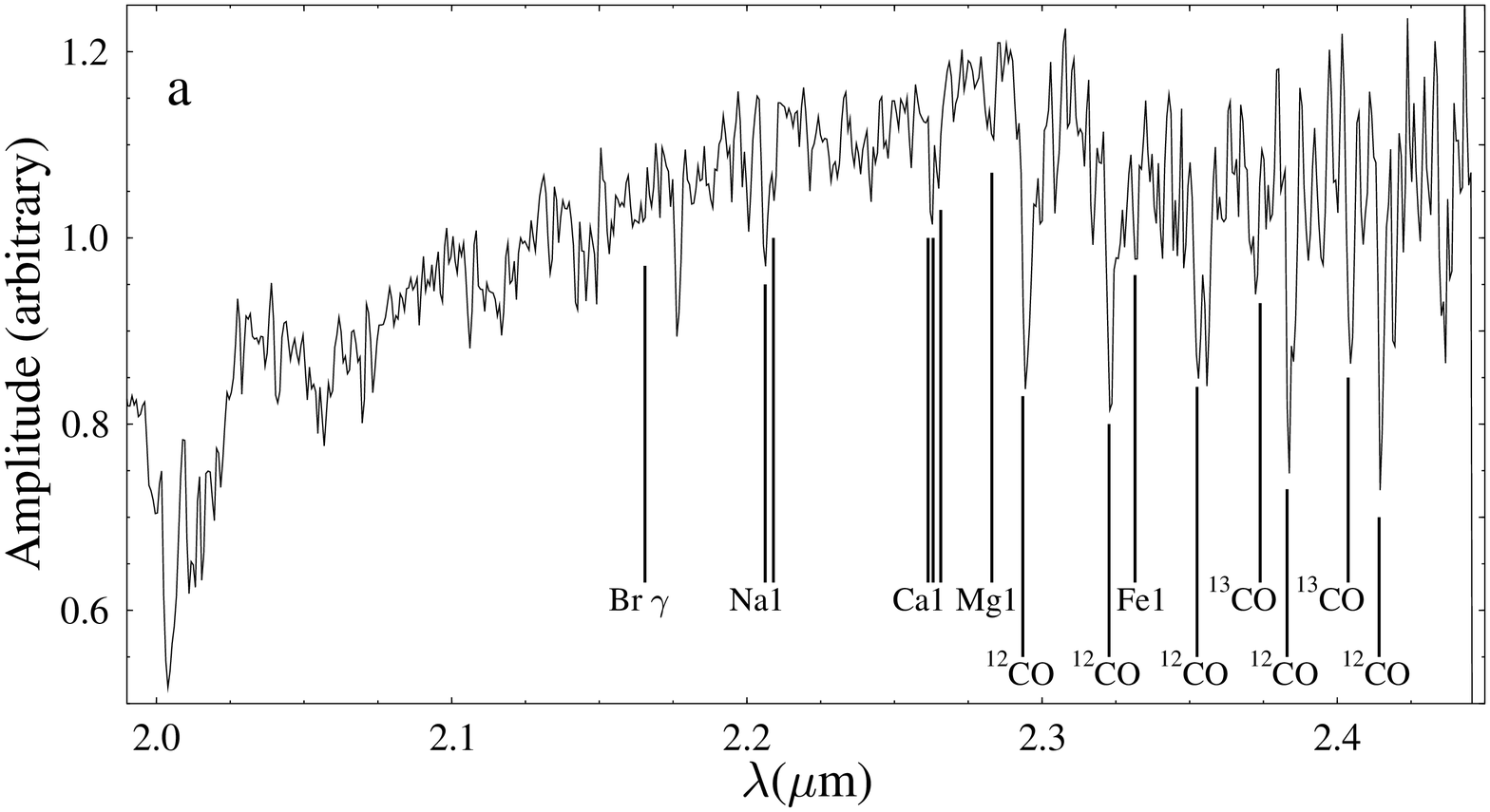}
  \hfill
  \includegraphics[height=.2\textheight]{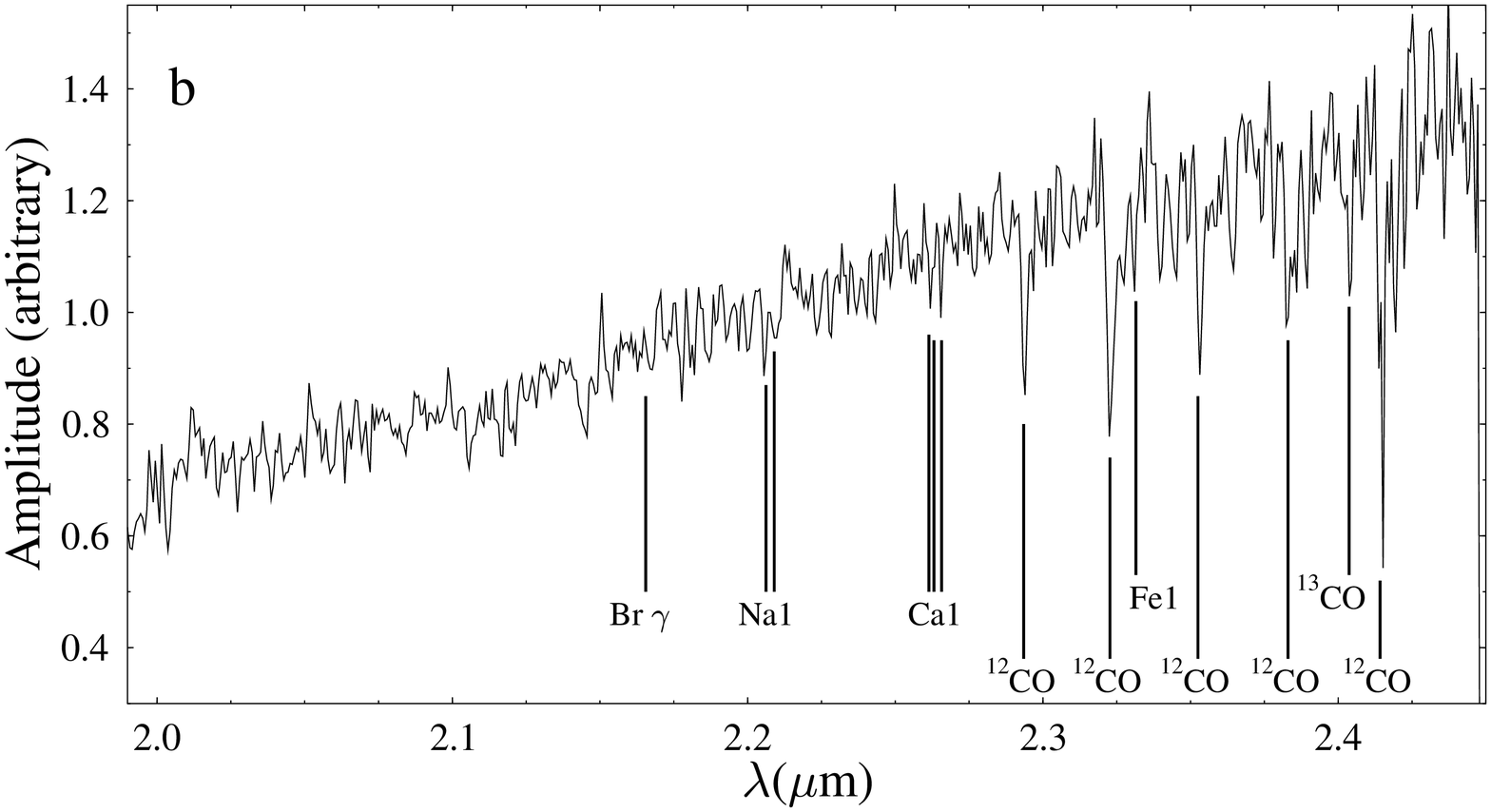}
  \caption{Examples of 2 of the $\ks$-band spectra obtained of possible
  counterparts to the Galactic centre X-ray sources.From the sample of
  36 targets for which spectra were obtained, none had obvious
  signatures of accretion ie: Brackett-$\gamma$ emission
  \citep{band97}. The spectra of almost all of the stars observed
  appeared to be K and M type giants with strong CO absorption
  features as well as other metal absorption lines.}
\label{2spec}
\end{figure}

Follow-up spectroscopic observations using VLT/ISAAC of 28 of these
targets, some with multiple candidate counterparts, yielded 36
spectra. The spectra of almost all the candidate counterparts were
late type red giant stars, identified by their strong CO absorption
bands beyond $2\,{\rm \mu m}$, and metal absorption lines. None
of these spectra contained evidence that the star was in an accreting
system; Brackett-$\gamma$ emission lines were absent from all (see
Fig. \ref{2spec}). We discuss a few of the possible reasons why the
accretion signatures were not seen in these spectroscopic targets.

The accretion signatures could be too weak to be measurable in the
spectra if the rate of accretion was low, or it could be self-absorbed
by a corresponding absorption feature of the companion star. Another
possibility is that the physical composition of the systems
means that there is no accretion at the time of observation;
an example is the propeller phase which inhibits accretion
onto neutron stars with strong magnetic fields.

More likely is that the candidate counterparts, for which we obtained
spectroscopy, were not the counterparts to the X-ray sources, but
merely coincidental astrometric matches due to the high stellar
density at the Galactic centre \citep{gosl06, band05}. Our imaging
survey obtained a limiting $\ks$ magnitude of 20. Assuming an average
extinction across the survey region (an assumption that is not valid,
but used in this case for an example, see Section: Extinction), the
survey will have been able to detect all stars on the giant branch as
well as all stars of A\,V main sequence or earlier. If the matches are
incidental, then the counterparts will be part of the later type main
sequence stars beyond the limits of this survey.

\section{X-ray source counterpart properties} 

\begin{figure}[h]
  \includegraphics[height=.3\textheight]{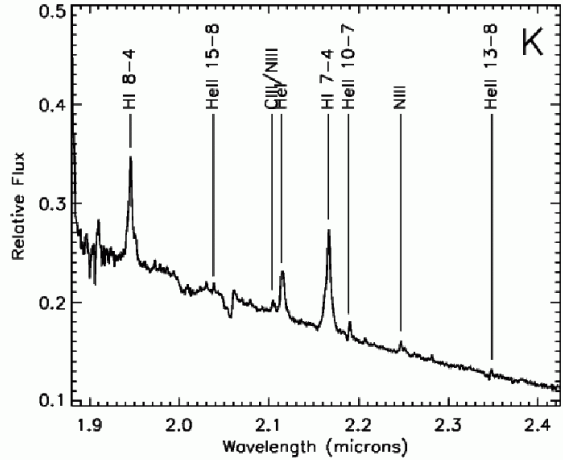}
  \caption{The $\ks$-band spectra of the first of the Galactic centre
  x-ray sources to have a spectroscopically identified
  counterpart. The emission line that identifies this target as an
  accreting system is the prominent Brackett-$\gamma$ (H1 7-4) line at
  $2.16\,{\rm \mu m}$ \citep{band97}. The system may also contain a
  wind or outflow as some of the lines show a P-Cygni profile. {\bf
  Image courtesy of Valerie Mikles, U. Florida} \citep{mikl06}}
\label{mikspec}
\end{figure}

Combining our results with those of a second group \citep[Mikles et
al., U. Florida; ][]{mikl06} who have spectroscopically confirmed one
counterpart to a {\it Chandra} Galactic centre X-ray source (see
Fig. \ref{mikspec}), we are able to draw some early conclusions as to
the probable population of X-ray sources at the Galactic centre.

A small proportion, $ \sim 2$-$3 \% $ may be high-mass X-ray binaries
or wind accreting neutron stars detectable in similar surveys. These
will most likely be the brighter sources in the population with $ L_X
\sim 10^{35} \, {\rm ergs^{-1}} $.

The vast majority of the population, those with lower luminosities, $
L_X \sim 10^{33} \, {\rm ergs^{-1}} $ will have main sequence
counterparts with lower masses are likely to be canonical low-mass
X-ray binaries with a main sequence counterpart or cataclysmic
variables. These counterparts will not be observable without very
deep, high resolution imaging.
 
\section{Extinction}

As part of our attempts to provide accurate photometry for the
candidate counterparts to the X-ray sources, we discovered that
infrared extinction towards the Galactic centre varies on scales much
smaller than previously observed, and that the absolute value of
extinction is also much greater than previously thought in some
regions. The extinction can rise to levels of as much as $A_K \sim 6$,
and varies on angular scales of $5\arcsec - 15\arcsec$, corresponding
to a physical scale of $0.2\,{\rm pc} - 0.6\,{\rm pc}$ at a Galactic
centre distance of $8.5\,{\rm kpc}$ \citep{gosl06}.

\begin{figure}[h]
\vspace{-50pt}
  \includegraphics[height=.6\textheight]{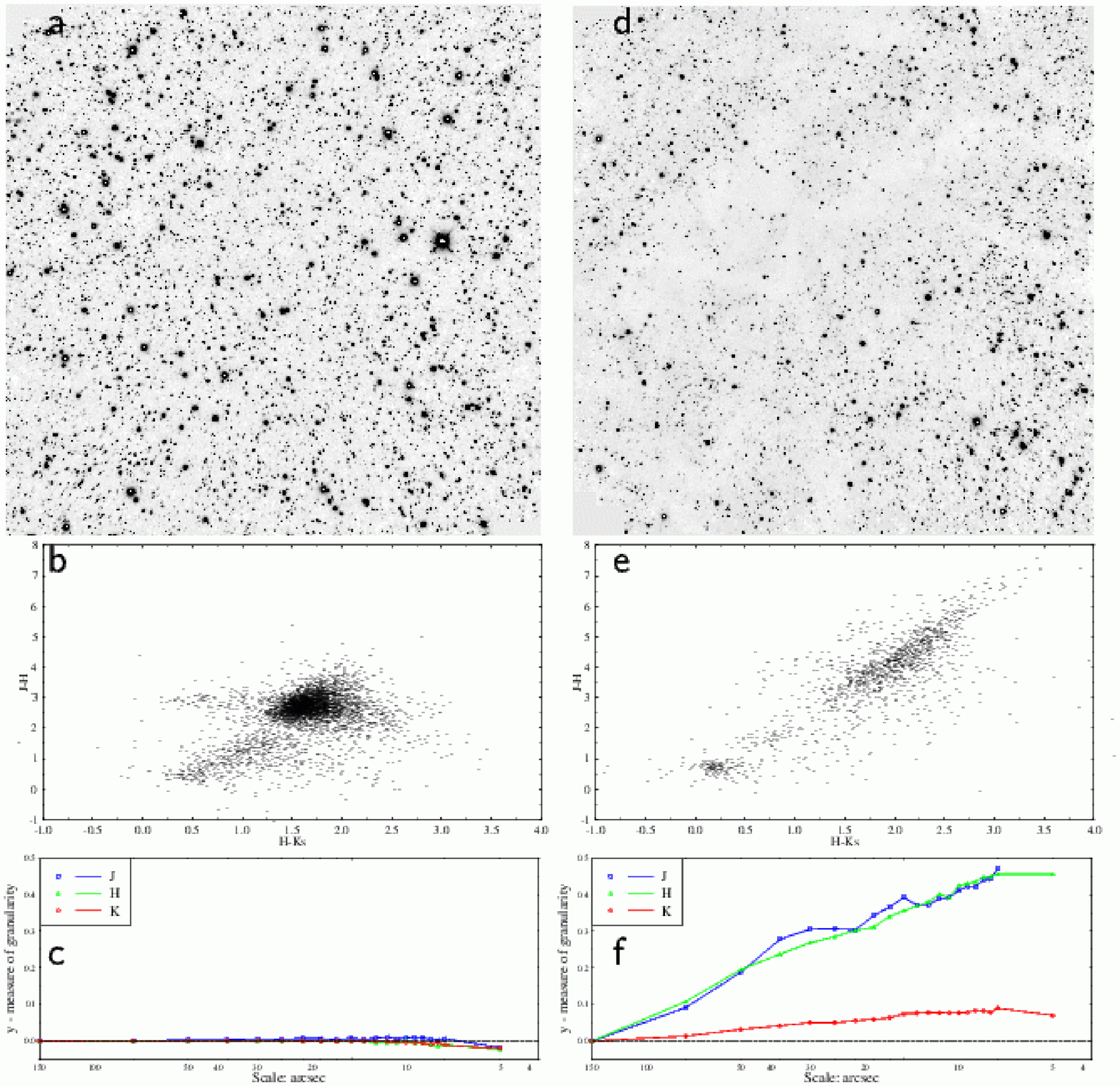}
  \caption{$(a)$ $\ks$ image of a field with no apparent structure in
  the stellar distribution. $(b)$ C-C diagram of the stars in the
  field shown in $(a)$. There are two main loci of stars: the local
  population to the bottom-left and the Galactic centre population in
  the centre. $(c)$ The measure of granularity for the field shown in
  $(a)$. In all three bands it does not deviate from zero, as expected
  for a random distribution. $(d)$ $\ks$ image of a field with obvious
  regions of low stellar density compared to the field average. $(e)$
  C-C diagram for the field shown in $(d)$. Note, compared to the C-C
  diagram in $(b)$, the locus for the Galactic centre stars is
  extended to high reddening. $(f)$ The measure of granularity for the
  field shown in $(d)$ shows that there is measurable structure in all
  three bands. {\bf Reproduced by permission of the AAS} \citep{gosl06}}
\end{figure}

These results are in agreement with other papers such as
\citet{schu99} who suggested that smaller-scale structures in the
extinction distribution, not measurable in their survey, were
responsible for the observed double-peaks in histograms of stellar
number versus $A_V$ in the Galactic centre. A recent paper by
\citet{nish06a} found that the infrared extinction varies from
sight-line to sight-line in the regions around the Galactic centre,
and that the previously stated universality of the infrared extinction
law for the Galactic centre is not valid.

We will produce a full and comprehensive map of the extinction for all
the fields observed as part of this survey (Gosling et al. {\it in
prep.}), and will work as part of the UKIDSS and VISTA survey team to
completely map the extinction towards the Bulge.

\section{Conclusions}

We have presented results from the search for the counterparts to the
X-ray sources of the Galactic centre. We found astrometric matches to
the positions of $\sim 70\%$ of the X-ray sources. The photometry of
these sources and follow up spectroscopy revealed that almost all of
these candidate counterparts were late type K and M giants. The
spectroscopy of this team, combined with that of the team of Mikles et
al. in Florida discovered evidence that only one of these candidates
could be positively identified as the counterpart to an X-ray
sources. From this we can estimate that $ \sim 2$-$3 \% $ may be
high-mass X-ray binaries or wind accreting neutron stars and that the
rest are low-mass X-ray binaries and cataclysmic variables.

\section{Acknowledgements}

AJG would like to thank the UK Particle Physics and Astronomy Research
Council for his studentship. KMB thanks the Royal Society for a
University Research Fellowship.  This work is based on observations
made with the ESO VLT at Paranal under imaging programme ID
071.D-0377(A) and spectroscopic programme ID 075.D-0361(A).

\end{document}